\documentclass[12pt]{amsart}
\usepackage{amsmath}
\usepackage{amsthm}
\usepackage{amsfonts}
\usepackage{graphicx}
\usepackage{amssymb} 
\newenvironment{nouppercase}{
  
  \renewcommand{\uppercasenonmath}[1]{}}{}

\begin{document}

\title[Generating self-similar membrane solutions]
{Generating self-similar membrane solutions}
\author[Jens Hoppe]{Jens Hoppe}
\address{Braunschweig University, Germany}
\email{jens.r.hoppe@gmail.com}

\begin{abstract}
Several ways to reduce to a first order ODE the non-linear PDE's governing
the relativistic motion of an axially symmetric membrane in 4 space time dimensions,
as well as examples for a previously found non-trivial transformation generating
solutions, are given
\end{abstract}

\begin{nouppercase}
\maketitle
\end{nouppercase}
\thispagestyle{empty}
\noindent
For several decades only very few non-trivial membrane solutions in 4 space-time dimensions (cp. \cite{1}) were known, in level-set form most notably 
\begin{equation}\label{eq1}
(t^2 +x^2 +y^2 -z^2)(t+z)^2 = C
\end{equation}
(for $C < 0$ genuinely time-like; non-compact, for each $t$ infinitely extended).\\
Recently \cite{2,3,4,5,6,7}, some new solutions, e.g. the `fast moving sharps drop' \cite{8} 
\begin{equation}\label{eq2}
(t^2 +x^2 +y^2 -z^2) - 6C\sqrt{x^2+y^2}(t+z)^2 + 3C^2(t+z)^4 = 0,
\end{equation}
were found, and methods proposed to generate infinitely many other solutions from them.\\
The simplest way to describe axially symmetric membranes (noted already in \cite{9}) arises from a light-cone description resulting in the Hamiltonian equations of motion
\begin{equation}\label{eq3}
\ddot{R} = R(RR')'
\end{equation}
for the radius $R(\tau, \mu)$ depending on (light-cone `time') $\tau := \frac{t+z}{2}$ and (spatial) parameter $\mu$ that is chosen such that the metric induced on the swept-out 3-manifold $\mathfrak{M}$ is diagonal, resp. that $\zeta(\tau, \mu) := t-z$ satisfies
\begin{equation}\label{eq4}
\zeta' = \dot{R}R', \; \dot{\zeta} = \frac{1}{2}(\dot{R}^2 + R^2R'^2),
\end{equation}
$\mathfrak{M}$ described parametrically by
\begin{equation}\label{eq5}
x^{\mu}(\tau, \mu, \psi) = 
\begin{pmatrix}
\tau + \zeta(\tau, \mu)/2\\[0.15cm] 
R(\tau, \mu)\left( 
\begin{smallmatrix} 
\cos \psi \\ \sin \psi
\end{smallmatrix}
\right)\\[0.15cm]
\tau - \zeta (\tau, \mu)/2
\end{pmatrix},
\end{equation}
and $\zeta$ having an intriguing partner $\kappa(\tau,\mu)$ satisfying
\begin{equation}\label{eq6}
\kappa' = \dot{\zeta}, \; \dot{\kappa} = R^2 \zeta'.
\end{equation}
The solutions of \eqref{eq3} corresponding to \eqref{eq1}$_{C=\frac{16\epsilon}{3}}$ and \eqref{eq2}$_{C=\frac{\epsilon}{4}}$ are, respectively, 
\begin{equation}\label{eq7}
R = \sqrt{2}\frac{\sqrt{\mu^2+\epsilon}}{\tau} \; \text{and} \; R = \sqrt{2}\frac{\mu}{\tau} + \epsilon \tau^2,
\end{equation}
the case $\epsilon = 0$ corresponding to the moving hyperboloids 
$z(t,x,y) = \pm \sqrt{t^2 + x^2 + y^2}$, resp.
\begin{equation}\label{eq8}
2\zeta\tau + R^2 = 0.
\end{equation}
In \cite{7} is was noted that when expanding $R = R_0 + \epsilon R_1 + \epsilon^2 R_2 + \ldots$ for solutions of \eqref{eq3} with $R_0 = \pm \sqrt{2} \frac{\mu}{\tau}$ the resulting (linear) equation for $R_1$ will be solved by
\begin{equation}\label{eq9}
R_1 = c\tau^{\alpha} \mu^{\beta}, \; \beta^2 + \beta +1 = \alpha (\frac{\alpha-1}{2}),
\end{equation}
\eqref{eq2} corresponding to $\alpha = 2,\, \beta = 0$ (and all $R_{n>1}$ chosen to be zero), while \eqref{eq1} corresponds to $\alpha = \beta = -1$,
\begin{equation}\label{eq10}
R = \sqrt{2}(\frac{\mu}{\tau} + \frac{\epsilon}{2\mu\tau}-\frac{\epsilon^2}{8\mu^3\tau}+ \ldots).
\end{equation}
In \cite{6} the following procedure was noted for obtaining a new solution $R^*(\tau,\mu)$ of \eqref{eq3}, from any given $R$ satisfying \eqref{eq3}: invert $\zeta(\tau, \mu)$, $\kappa(\tau, \mu)$ (satisfying \eqref{eq4} and \eqref{eq6}) to obtain $\tau(\zeta, \kappa)$, $\mu(\zeta, \kappa)$ (note that the Jacobian $\dot{\zeta}\kappa' - \zeta'\dot{\kappa}$, due to \eqref{eq4}/\eqref{eq6}, equals $\mathcal{L}^2 = \frac{1}{4}(\dot{R}^2 - R^2R'^2)$, i.e. is non-vanishing as long as the Lagrangian for \eqref{eq3} is different from zero); $R^*(\tau, \mu)$ is then defined by $R(\tau,\mu) = R^*(\zeta,\kappa)$.\\
By straightforward (though somewhat tedious) calculations one can perturbatively deduce that (for small $\delta$)   
\begin{equation}\label{eq11}
R(\tau,\mu) = \pm \sqrt{2} (\frac{\mu}{\tau} + \delta \tau^{\alpha}\mu^{\beta} + O(\delta^2))
\end{equation}
implies
\begin{equation}\label{eq12}
R^*(\tau, \mu) = \mp\sqrt{2}(\frac{\mu}{\tau} + \delta^* \tau^{\alpha^*}\mu^{\beta^*} + O(\delta^2))
\end{equation}
with
\begin{equation}\label{eq13}
\begin{split}
\alpha^* & = -3\alpha - 4\beta, \qquad \beta^* = 2\alpha + 3\beta \\
\delta^* & = \delta(-)^{\alpha}\bigg(2\frac{\alpha-\beta}{\beta +1}(1+\frac{\alpha-2}{\beta+2}) -1\bigg);
\end{split}
\end{equation}
while $\alpha^{**} = \alpha$, $\beta^{**} = \beta$ is immediate,the calculation of $\delta^*$ (and $\delta^{**} = \delta$) is more involved.\\
From 
\begin{equation}\label{eq14}
\begin{split}
\zeta  & = -\frac{\mu^2}{\tau^3} + 2\delta' \tau^{\alpha-2} \mu^{\beta+1} + O(\delta^2)\\
\kappa  & = \frac{\mu^3}{\tau^4} + 2\delta' \frac{\alpha-2}{\beta+2} \tau^{\alpha-3} \mu^{\beta+2} + O(\delta^2)
\end{split}
\end{equation}
\begin{equation}\label{eq15}
\delta' = \delta\frac{(\alpha - \beta)}{\beta + 1}
\end{equation}
(note that, needed for verifying \eqref{eq4}/\eqref{eq6}, \eqref{eq9} implies $\frac{\alpha-2}{\beta+1}(\alpha-\beta) = 2-\alpha+2\beta$, $(\alpha-\beta)\frac{\alpha-2}{\beta+1}\,\frac{\alpha-3}{\beta+2} = 2\alpha - 2\beta -4$, assuming $\beta \neq -1,-2$; for the important case, $\beta = -1 = \alpha$, \eqref{eq14} still holds if then using $\delta' = -\frac{\delta}{3}$), and 
\begin{equation}\label{eq16}
\tau = -\frac{\kappa^2}{\zeta^3}(1+\delta\tau), \quad \mu = \frac{\kappa^3}{\zeta^4}(1+\delta\mu),
\end{equation}
one obtains
\begin{equation}\label{eq17}
\begin{split}
(1+\delta\tau)^3 & = (1+\delta\mu)^2 - 2\hat{\delta}(-)^{\alpha+1}(1+\delta\tau)^{\alpha+1}(1+\delta\mu)^{\beta+1}\\[0.15cm]
(1+\delta\tau)^4 & = (1+\delta\mu)^3 + 2\hat{\delta}(-)^{\alpha+1}\frac{\alpha-2}{\beta+2}(1+\delta\tau)^{\alpha+1}(1+\delta\mu)^{\beta+1}\\[0.15cm]
\hat{\delta} & = \delta'\frac{\kappa^{2\alpha + 3\beta-1}}{\zeta^{3\alpha+4\beta-1}}, 
\end{split}
\end{equation}
which for the leading order terms implies 
$
\left( 
\begin{smallmatrix}
3 & -2 \\ 4 & -3
\end{smallmatrix}
\right) 
{\tau_1 \choose \mu_1} = 2\delta'(-1)^{\alpha+1}{-1 \choose \frac{\alpha-2}{\beta+2}}$ i.e.
\begin{equation}\label{eq18}
\begin{pmatrix}
\tau_1 \\ \mu_1
\end{pmatrix} = 2 \delta' (-)^{\alpha}
\begin{pmatrix}
-3 & 2 \\ -4 & 3
\end{pmatrix}
\begin{pmatrix}
-1 \\ \frac{\alpha-2}{\beta+2}
\end{pmatrix}.
\end{equation}
Insertion of \eqref{eq16} into $R(\tau,\mu)$ on the other hand gives \eqref{eq12}, with, in leading order,
\begin{equation}\label{eq19}
\delta^* = \mu_1 - \tau_1 - \delta(-)^{\alpha}.
\end{equation}
Note that \eqref{eq18} may still be used for the singular choice $\beta = -1 = \alpha$ (with $\delta' = -\frac{\delta}{3}$ then) which will now be discussed explicitly ($\delta = \frac{\epsilon}{2}$ i.e. $\delta' = -\frac{\epsilon}{6}$, to agree with \eqref{eq1}): $\alpha = \beta = -1$, $\alpha^* = 7$, $\beta^* = -5$  
\begin{equation}\label{eq20}
\begin{split}
\begin{pmatrix}
\tau_1 \\ \mu_1
\end{pmatrix} & = -2\cdot (-\frac{\epsilon}{6})
\begin{pmatrix}
-3 & 2 \\ -4 & 3 
\end{pmatrix}
\begin{pmatrix}
-1 \\ -3
\end{pmatrix} = \epsilon 
\begin{pmatrix}
-1 \\ -\frac{5}{3}
\end{pmatrix} \\
\delta^* & = \epsilon (-\frac{5}{3} + 1 + \frac{1}{2}) = -\frac{\epsilon}{6}\\
R^{*2} + 2\tau \zeta^* & = 4 \delta^* \tau^{\alpha^* -1} (-\zeta^*\tau^3)^{\frac{\beta^*+1}{2}}(1+\frac{\alpha^*-\beta^*}{\beta^*+1})+ \ldots = \frac{4}{3}(\zeta^*)^{-2}\epsilon, 
\end{split}
\end{equation}
consistent with the involutive transformation $R\rightarrow R^*$ corresponding to $\tau \leftrightarrow \zeta$ (which /note the unsymmetrically distributed factors of $\sqrt{2}$; in the string literature one conventionally takes $x_{\pm} = \frac{1}{\sqrt{2}}(t \pm z)$/ is a combination of the reflection $z\rightarrow -z$ and a boost with rapidity $\gamma$,  $\sinh \gamma = \frac{3}{4}$). $\tau_1^*$, $\mu_1^*$ are given (``again'') by \eqref{eq18}$_{\alpha^* = 7, \beta^* = -5}$
\begin{equation}\label{eq21}
\begin{split}
\begin{pmatrix}
\tau_1^* \\ \mu_1^*
\end{pmatrix} & = 2{\delta^*}' (-)^{\alpha^*}  
\begin{pmatrix}
-3 & 2 \\ -4 & 3
\end{pmatrix}
\begin{pmatrix}
-1 \\ \frac{\alpha^* -2}{\beta^* + 2} \end{pmatrix} \\[0.15cm]
& = -2(\frac{7+5}{-4}) \begin{pmatrix}
-3 & 2 \\ -4 & 3
\end{pmatrix}
\begin{pmatrix}
-1 \\ -\frac{5}{3}
\end{pmatrix} (-\frac{\epsilon}{6}),
\end{split}
\end{equation}
hence
\begin{equation}\label{eq22}
\delta^{**} = \epsilon (1 - \frac{1}{3}-\frac{1}{6}) = \frac{\epsilon}{2} = \delta.
\end{equation}
Generally, 
\begin{equation}\label{eq23}
\frac{\delta^{**}}{\delta}  = (1 + 2\frac{\alpha-\beta}{(\beta+1)(\beta+2)})(1+2\frac{\alpha^*-\beta^*}{(\beta^*+1)(\beta^*+2)}).
\end{equation}
While in the perturbative expansion \eqref{eq11} $\alpha$ and $\beta$ are assumed to be related to each other, i.e. satisfy \eqref{eq9}, it is interesting to note that
\begin{equation}\label{eq24}
R(\tau, \mu) = \frac{\mu}{\tau} f (x := \tau^{\alpha'}\mu^{\beta'})
\end{equation}
consistently reduces \eqref{eq3} to an ODE for $f$,
\begin{equation}\label{eq25}
\begin{split}
2(f-\alpha'x f') & - \alpha'x(f-\alpha'xf')' \\
& = \beta' x f^2 (f+\beta x f')' + f(f + \beta' x f')^2,
\end{split}
\end{equation}
for \textit{any} $\alpha'$, $\beta' \in \mathbb{R}$.\\\\
$f(x) = g(u := \ln x)$ gives $xf' = g'$, $x^2f'' = g''-g'$, hence
\begin{equation}\label{eq26}
2g - 3\alpha' g' + \alpha'^2 g'' = g''g^2\beta'^2 + 3\beta'g'g^2 + \beta'^2 g g'^2 + g^3
\end{equation}
resp. (setting $g' = G(g)$, implying $g'' = GG'$)
\begin{equation}\label{eq27}
GG'(\alpha'^2 - \beta'^2g^2) = \beta'^2 g G^2 + 3(\beta'g^2 + \alpha')G + g^3-2g
\end{equation}
resp. (standard Abel's ODE of the second kind)
\begin{equation}\label{eq28}
GG' = f_2(g)G^2 + f_1(g)G + f_0(g),
\end{equation}
which upon letting $w(g) := \sqrt{\alpha'^2 - \beta'^2 g^2}G$ brings \eqref{eq28} into canonical form,
\begin{equation}\label{eq29}
ww' = F_1w + F_0 \qquad F_1 = \frac{3(\alpha'+\beta'g^2)}{\sqrt{\alpha'^2-\beta'^2g^2}}, \; F_0 = g(g^2 -2).
\end{equation}
\begin{equation}\label{eq30}
\begin{split}
\int \sqrt{1-\dot{r}^2 + r'^2} \,rdzdt & \:\,= \int \sqrt{1 - 2\dot{s}s'}\,s d\tau d\zeta \\
& \: \,= \int \sqrt{\tau_R^2 + 2\tau_{\zeta}} \,RdRd\zeta \\
& \overset{\text{resp.}}{=} \int \sqrt{\zeta'^2 + 2\dot{\zeta}} \,RdRd\tau 
\end{split}
\end{equation}
gives (cp.\cite{10})
\begin{equation}\label{eq31}
s(\ddot{s}s'^2 + s''\dot{s}^2 - 2\dot{s}s'\dot{s}') + 2(s \dot{s}' - \dot{s}s') + 1 = 0
\end{equation}
for $R = s(\tau, \zeta)$, and for $\tau(\zeta, R)$
\begin{equation}\label{eq32}
\ddot{\tau} - 2\dot{\tau}\tau'' + 2 \tau'\dot{\tau}' = \frac{\tau'}{R} (\tau'^2 + 2\dot{\tau}),
\end{equation}
resp.
\begin{equation}\label{eq33}
\ddot{\zeta} - 2\dot{\zeta}\zeta'' + 2 \zeta'\dot{\zeta}' = \frac{\zeta'}{R} (\zeta'^2 + 2\dot{\zeta})
\end{equation}
for $\zeta(\tau,R)$ (implying that $\tilde{\zeta} = \alpha\zeta(\alpha\gamma^2\tau, \gamma R)$ is a solution if $\zeta$ is).\\\\
Note that \eqref{eq32} can also be derived from \eqref{eq4} by the hodograph transformation $(\tau,\mu)\leftrightarrow R, \zeta$ giving 
\begin{equation}\label{eq34}
\begin{split}
\mu_{\zeta}(\tau^2_R + \tau_{\zeta}) & = \mu_R \tau_{\zeta} \tau_R \\
\mu^2_{\zeta} + 2\mu_{\zeta}\mu_R\tau_R + R^2\tau_{\zeta}^2 & = 2\mu^2_R\tau_{\zeta}
\end{split}
\end{equation}
which implies
\begin{equation}\label{eq35}
\pm \dot{\mu} = \frac{R\dot{\tau} \tau'}{\sqrt{\tau'^2 + 2\dot{\tau}}}, \qquad \pm \mu'= \frac{R(\dot{\tau}+\tau'^2)}{\sqrt{\tau'^2 + 2\dot{\tau}}},
\end{equation}
hence \eqref{eq32}.
If $s$ solves \eqref{eq31}, which is also invariant under $\tau \leftrightarrow \zeta$, so does $\tilde{s}(\tau, \zeta) = es(a\tau,b\zeta)$ if
\begin{equation}\label{eq36}
e^2ab =1.
\end{equation}
\eqref{eq32} has rational solution
\begin{equation}\label{eq37}
\tau = \frac{c}{2\zeta^3} - \frac{R^2}{2\zeta} = \zeta^{-3}(\frac{c}{2} - \frac{R^2}{2}\zeta^2)
\end{equation} 
\begin{equation}\label{eq38}
\tau = \frac{c}{2}\zeta^5 + \frac{R^2}{2\zeta} = \zeta^5(\frac{c}{2} + \frac{R^2}{\zeta^6})
\end{equation}
while 
\begin{equation}\label{eq39}
\tau = \zeta^A T(\zeta^B R =: q),\quad  A+2B+1 = 0
\end{equation}
gives 
\begin{equation}\label{eq40}
\begin{split}
A(A-1)T & + B(2A+ B-1)T'q + B^2q^2 T''\\[0.15cm]
& = 2A(T''T - T'^2) + \frac{T'^3}{q} + 2\frac{T'AT}{q},
\end{split}
\end{equation}
with simple solutions $T = \frac{c}{2}-\frac{q^2}{2}$ (for $A = -3,\, B=1$) and $T = \frac{c}{2} + \frac{q^2}{2}$ (for $A=5,\, B=-3$). 
\begin{equation}\label{eq41}
\zeta(\tau, R) = -\frac{R^2}{2\tau} h(\tau^aR^b =: z)
\end{equation} 
on the other hand, for any $a,b \in \mathbb{R}$, gives
\begin{equation}\label{eq42}
\begin{split}
[az(h - azh')' & - 2(h-azh')]  \\
& + (h - azh')[(2h + bzh') + bz(2h + bzh')']\\
& - (2h + bzh')[2(h -azh')+ bz(h-azh')']\\
& + (2h + bzh')[\frac{(2h+bzh')^2}{4} + (h-azh')] \\
& = 0,
\end{split}
\end{equation}
using  \; $\dot{•} = \partial_{\tau}= \frac{az}{\tau}\partial_z$, \; $' = \partial_R = \frac{bz}{R}\partial_z$.
Due to the otherwise unmatched powers arising from $(2h + bzh')^3$ one can try
\begin{equation}\label{eq43}
h(z) = 1+ dz +ez^2 \;\; \text{for}\; b=-1,
\end{equation} 
giving (from $O(z)$ and $O(z^3)$)
\begin{equation}\label{eq44}
d(a^2-a-2) = 0, \quad d(\frac{d^2}{4} + e(1-2a)) = 0
\end{equation} 
and
\begin{equation}\label{eq45}
2e(1-2a)a + d^2 + 2(\frac{d^2}{4} + e(1-2a)) = 0
\end{equation} 
in  $O(z^2)$, which for $d \neq 0$ implies $a=2$, $e = \frac{d^2}{12}$ (corresponding to \eqref{eq2}) while for $d=0$, $2e(1-2a)(a+1) = 0$ (i.e., corresponding to \eqref{eq1}, $a=-1$; or, corresponding to a simple constant shift in $\zeta$, $a= \frac{1}{2}$).\\
For 
\begin{equation}\label{eq46}
h(z) = -1 + dz +ez^2, \; b= -1
\end{equation}
on the other hand one gets (from $O(z, z^2, z^3)$) 
\begin{equation}\label{eq47}
d(a^2 -5a + 1) = 0, \; \frac{d^2}{4} = e(1-2a)(\frac{a-3}{3}), \; (\frac{d^2}{4} + e(1-2a))d = 0
\end{equation}
which implies $d=0$ and $a=3$, corresponding to 
\begin{equation}\label{eq48}
\zeta = \frac{c}{2} \tau^5 + \frac{R^2}{2\tau} = -\frac{R^2}{2\tau}(-1-\frac{c\tau^6}{R^2}) 
\end{equation}
or $ d= 0, \, a = \frac{1}{2}$, $\zeta = -\frac{R^2}{2\tau}(-1 + \frac{c\tau}{R^2}),$ 
i.e. shifting $\zeta = \frac{R^2}{2\tau}$ (which does solve \eqref{eq33}, but does not define a minimal 3-manifold, due to $(\zeta'^2 + 2\dot{\zeta})$, cp.\eqref{eq30}, vanishing identically in that case) by a constant.\\\\
Corresponding to eq.\eqref{eq35} of \cite{6} (with $\frac{\dot{F}}{F} = f$), on the other hand,
\begin{equation}\label{eq49}
\zeta = \frac{R^2}{2}\cdot f(\tau)
\end{equation}
gives
\begin{equation}\label{eq50}
\ddot{f} = 2 f^3, \quad \dot{f}^2 = f^4+\delta
\end{equation}
(note that if $\dot{F}^2 = F^4 +1$, as in \eqref{eq36} of \cite{6}, i.e. $F$ a hyperbolic lemniscate-function, $\frac{\dot{F}}{F}$ indeed also satisfies \eqref{eq50}, but with $\delta = -4$). \eqref{eq49}/\eqref{eq50} is a special case of the solutions found in \cite{10},
\begin{equation}\label{eq51}
\zeta = \frac{R^2}{2}D(\tau) - C_1 - C_2 \int e^{\int 4D}, \quad \ddot{D} = 2D^3;
\end{equation}
while $C_1$ is trivial,
the existence of the additional free constant $C_2$, having to do with the non-linear PDE \eqref{eq33} giving a \textit{linear} equation,
\begin{equation}\label{eq52}
\ddot{\zeta}_1 = 4D\dot{\zeta}_1,
\end{equation}
if inserting $\zeta = \frac{R^2}{2}D + \zeta_1(\tau)$ with $\zeta_1$ independent of $R$, is non-trivial.\\\\
From \eqref{eq31} one can easily see that $g:=\frac{1}{2}s^2$ satisfies
\begin{equation}\label{eq53}
(\ddot{g}g'^2 + g''\dot{g}^2 - 2\dot{g}'\dot{g}g') + 4(g\dot{g}' - \dot{g}g') + 2g = 0,
\end{equation}
while $g = \pm \tau \zeta + f$ gives
\begin{equation}\label{eq54}
\begin{split}
(\ddot{f}f'^2 & + f''\dot{f}^2 - 2\dot{f}f'\dot{f}')  \\
& \pm 2 \big( \ddot{f}\tau f' + f'' \zeta \dot{f} - \dot{f}f' - (\zeta f' + \tau \dot{f})\dot{f}' + 2(f \dot{f}' - \dot{f}f') \big)\\
& + \ddot{f}\tau^2 + f'' \zeta^2 - 2 (\dot{f}' \tau \zeta + f'\zeta + \dot{f}\tau)\\
& \pm 4(f + \tau \zeta \dot{f}' - \zeta f' - \tau\dot{f} \pm \frac{1}{2}f)\\
& = 0,
\end{split}
\end{equation}
the linear part $L_{\pm}[f]$ vanishing for all $f = \tau^a\zeta^b$ with $a^2+b^2 + 2ab(\pm 2 -1) + (a+b)(-3\mp 4)+ 2(1\pm 2) = 0$ while (for that Ansatz) the quadratic and cubic parts together vanish only if either $a$ or $b$ equals zero; so, while in general $f=\delta f_1 + \delta^2 f_2 + \ldots$ in each order $\delta^n$ gives $L_{\pm}[f_n] = I_n^{\pm}(f_{m<n})$ one may choose all $f_{n>1} \equiv 0$ if $f_1 = \tau^a$ (or $\zeta^a$) with $a^2 + a(-3 \mp 4)+ 2(1\pm 2) = 0$.\\
Finally note that \eqref{eq11}/\eqref{eq14} imply
\begin{equation}\label{eq55}
R^2+2\zeta \tau = R^{\beta+1} \tau^{\alpha+\beta} 2 ^{\frac{3-\beta}{2}}(\delta + \delta') + \ldots
\end{equation}
(for $\alpha = \beta = -1$, $\delta = \frac{\epsilon}{2}$, $\delta' = \frac{-\delta}{3} = \frac{-\epsilon}{6}$; resp. $\alpha = 2, \, \beta  = 0, \, \delta = \frac{\epsilon}{\sqrt{2}}, \, \delta' = 2\delta$ recovering \eqref{eq1}$_{C = \frac{16\epsilon}{3}}$ and \eqref{eq2}$_{C=\frac{\epsilon}{4}}$ to leading order), and the following consequences of the ubiquitous first order equations
\begin{equation}\label{eq56}
Y' = \dot{X}, \quad \dot{Y} = R^2 X'
\end{equation}
that hold for $(X, Y) = (t, \varphi), (z,v), (\tau, \mu), (\tau, \kappa)$, with $R$ being the radius of the axially symmetric membrane (in the correspondingly relevant variables); just as in \cite{6} it was proven that $R^*(\zeta, \kappa) := R(\tau, \mu)$ satisfies \eqref{eq3} (with respect to $(\zeta, \kappa)$), it is e.g. also true that
\begin{equation}\label{eq57}
\ddot{\zeta} = (R^2 \zeta')'
\end{equation}
holds both in $(\tau, \mu)$ variables ($\zeta(\tau,\mu)$ defined by \eqref{eq4}) \textit{and} with respect to $(t, \varphi)$, using $\zeta = t- z(t,\varphi)$ (in fact, \textit{both} parts satisfy \eqref{eq57} separately, with respect to $(t, \varphi)$).
\begin{equation}\label{eq58}
\begin{array}{lll}
t =\tau +\frac{\zeta}{2}, & \varphi = \mu + \frac{\kappa}{2}  \\[0.15cm]
\tau  = \frac{1}{2}(t+z), & \mu = \frac{1}{2}(\varphi + v), & \kappa = \varphi - v\\[0.15cm]
v' = \dot{z}, & \dot{v} = r^2z'  
\end{array} 
\end{equation}
(implying $\ddot{z} = (r^2z')'$).\\
For the moving hyperboloids in orthonormal parametrization, a rare example of explicitly solving
\begin{equation}\label{eq59}
\dot{r}r' + \dot{z}z' = 0, \quad \dot{r}^2 + \dot{z}^2 + r^2(r'^2 + z'^2) = 1,
\end{equation}
one has
\begin{equation}\label{eq60}
\begin{array}{ll}
z = t\tilde{s}(u), & \tilde{s}(u) := \frac{\pm 1}{\sqrt{2}}(\sqrt{1+8u}+1)^\frac{1}{2} \\[0.15cm]
r^2 = t^2(s^2-1), & \tilde{s}^2(\tilde{s}^2-1) = 2u := \frac{2\varphi^2}{t^4}\\[0.15cm]
v = \frac{\varphi}{\tilde{s}}, & 2u\tilde{s}'(2\tilde{s}^2 - 1) = \tilde{s}^3 - \tilde{s}\\[0.15cm]
\tau = t(\frac{1+\tilde{s}}{2}) = t f(u), & \mu = \frac{1}{2}\varphi(1+ \frac{1}{\tilde{s}}) = \varphi g(u)\\[0.15cm]	\tilde{s}\tilde{s}'(2\tilde{s}^2 -1) = 1, & f-4uf' = g + 2ug' \\[0.15cm]
(\tilde{s}^2 - 1)f' = -2ug'.  
\end{array} 
\end{equation}
Great care is needed with field-dependent reparametrisations on the Lagrangian level, as naively 
\begin{equation}\label{eq61}
\int (\dot{R}^2 - R^2R'^2) d\tau d\mu = \int (\dot{r}^2 - r^2 r'^2) dt d \varphi,
\end{equation}
if only using $\dot{R} = \partial_{\tau}R = \dot{t}\dot{r} + \dot{\varphi}r'$, $R' = \partial_{\mu}R = t'\dot{r} + \varphi' r'$ and $\mu' = \dot{\tau}$, $\dot{\mu} = r^2 \tau'$ resp. $\varphi' = \dot{t}$, $\dot{\varphi} = R^2 t'$; but $r(t,\varphi)$ satisfies
\begin{equation}\label{eq62}
\ddot{r} - r^2 r'' = r(r'^2 + z'^2),
\end{equation}
rather than \eqref{eq3} (the last term in \eqref{eq62} naively \textit{not} arising from the RHS of \eqref{eq61}).\\
On the other hand, treating the radius $R$ as given (`external'),
\begin{equation}\label{eq63}
\frac{1}{2}\int (\dot{Z}^2 - R^2Z'^2) dxdy = \frac{1}{2} \int (\dot{\tilde{Z}}^2 - \tilde{R}^2 {{{}\tilde{Z}}}'{}^2) d\tilde{x}d\tilde{y} 
\end{equation}
proves that if 
\begin{equation}\label{eq64}
\ddot{Z} = (R^2 Z')',
\end{equation}
it will also hold in any other variables satisfying \eqref{eq56}; explaining e.g. why $\zeta(\tau, \mu) = t - z(t,\varphi) = \tilde{\zeta}(t, \varphi)$ satisfies \eqref{eq64} with respect to both $(\tau, \mu)$ as well as $(t,\varphi)$ variables (in fact: all pairs of variables $(\tilde{x}, \tilde{y}) = (X, Y)$ satisfying \eqref{eq56}; so in particular also $(z,v)$).\\\\
One more observation: the Ansatz
\begin{equation}\label{eq65}
r(t,\varphi) = t s(u := \frac{\varphi^2}{t^4}), \quad z(t, \varphi) = t\tilde{s}(u)
\end{equation}
not only gives (cp.\eqref{eq60})
\begin{equation}\label{eq66}
s(u) = \frac{1}{\sqrt{2}}(\sqrt{1+8u}-1)^\frac{1}{2}, \quad \tilde{s}^2 = s^2+1
\end{equation}
as a solution.\\ 
The equations
\begin{equation}\label{eq67}
\begin{array}{l}
s'(s-4us')+ \tilde{s}'(\tilde{s} - 4u\tilde{s}') = 0\\[0.15cm]
(s-4us')^2 + (\tilde{s}-4u\tilde{s}')^2 + 4us^2(s'^2 + \tilde{s}'^2) = 1
\end{array}
\end{equation}
also do have other solution -- which correspond to the freedom when inserting $r = zg(\frac{t}{z} =: x)$ into (cp.\eqref{eq30})
\begin{equation}\label{eq68}
\ddot{r}(1+r'^2) - r''(1-\dot{r}^2) - 2\dot{r}r'\dot{r}' + \frac{1-\dot{r}^2 + r'^2}{r} = 0,
\end{equation}
which gives
\begin{equation}\label{eq69}
gg''(1+g^2-x^2) - g'^2 + (g-xg')^2 + 1 =0
\end{equation}
resp. (having a host of solutions -- though difficult to obtain explicity)
\begin{equation}\label{eq70}
(2hh'' - h'^2)(1+h-x^2)-h'^2+4h+(2h-xh')^2 = 0
\end{equation}
for $h=g^2$.\\
Note that while $h = x^2-1$ is unphysical/singular (the factor $1-\dot{r}^2-r'^2$, that in the derivation of \eqref{eq68} was assumed to be non-vanishing, being identically zero in that case), $h = 1-x^2$ corresponds to \eqref{eq66} (a solution that helps to understand a subtle difficulty of eq.\eqref{eq28} of \cite{5}: the 2 signs there do \textit{not} correspond to 2 different `sectors', but -- in the derivation implicity having assumed $p^2-q^2$ to be non-zero -- have to be chosen according to whether $\dot{z}^2$ is bigger, or smaller than $r^2z'^2$; and inserting \eqref{eq65} into that equation, $D_{\pm}[r] = 0$ gives an interesting second-order ODE for $s(u) = P(y:=\frac{\varphi}{t^2}), \, \big( P''(P^2-4y^2)-2yP' + PP'^2 \big)P = p- \sqrt{p^2-q^2}$, $2p = 1-(P -2yP')^2 - P^2P'^2$, $q = PP'(P-2yP')$, of which $P^2(P^2 +1) = 2y^2$, $P' = \frac{2y}{P(2P^2+1)}$, $\tilde{s}^2 = P^2+1$, is only the simplest solution,-- which can equally be derived from \eqref{eq67}).\\\\
\textbf{Acknowledgement}:
I would like to thank V.Bach, J.Eggers, J.Fr\"{o}hlich, G.M.Graf, G.Huisken and T.Turgut for discussions.

\end{document}